
\magnification=\magstep1
\font\fvmib=cmmib10 scaled 500
\font\sixmib=cmmib10 scaled 600
\font\svnmib=cmmib10 scaled 700

\font\eitmib=cmmib10 scaled 800
\font\tenmib=cmmib10
\font\twlvmib=cmmib10 scaled \magstep1
\newfam\mibfam \def\mib{\fam\mibfam\tenmib}
\textfont\mibfam=\tenmib \scriptfont\mibfam=\svnmib
\scriptscriptfont\mibfam=\fvmib
\newfam\Mibfam 
\textfont\Mibfam=\twlvmib \scriptfont\Mibfam=\eitmib
\scriptscriptfont\Mibfam=\sixmib

\font\twlvrm=cmr12

\font\eitrm=cmr8

\font\sixrm=cmr6

\newfam\Rmfam \def\Rm{\fam\Rmfam\twlvrm}
\textfont\Rmfam=\twlvrm \scriptfont\Rmfam=\eitrm
\scriptscriptfont\Rmfam=\sixrm

\font\tenex=cmex10

\font\twlvbf=cmbx12
\font\eitbf=cmbx8
\font\sixbf=cmbx6
\newfam\Bffam \def\Bf{\fam\Bffam\twlvbf}
\textfont\Bffam=\twlvbf \scriptfont\Bffam=\eitbf
\scriptscriptfont\Bffam=\sixbf
\font\twlvmi=cmmi12
\font\eitmi=cmmi8
\font\sixmi=cmmi6

\newfam\Mifam 
\textfont\Mifam=\twlvmi \scriptfont\Mifam=\eitmi
\scriptscriptfont\Mifam=\sixmi

\font\twlvsy=cmsy10 scaled \magstep1

\font\eitsy=cmsy8

\font\sixsy=cmsy6

\newfam\Syfam 
\textfont\Syfam=\twlvsy \scriptfont\Syfam=\eitsy
\scriptscriptfont\Syfam=\sixsy





\def\references {\vfill\eject {\leftline {\Bf References}}\medskip}
\def\endreferences{\vfill\eject}
\gdef\jrnl#1#2#3#4{,~#1 {\bf ~#2}(#3),~#4}
\gdef\refto#1{$^{#1)}$}

\gdef\refis#1{\item{#1)\ }}
\def\t{\tilde}
\def\hb{\hfill\break}
\def\dl{\displaylines}
\def\ds{\displaystyle}
\def\eq{\eqalign}

\def\cosx{\cos k_{x}a}
\def\cosy{\cos k_{y}a}
\def\cosxd{\cos k'_{x}a}
\def\cosyd{\cos k'_{y}a}
\def\sinx{\sin k_{x}a}
\def\siny{\sin k_{y}a}
\def\sinxd{\sin k'_{x}a}
\def\sinyd{\sin k'_{y}a}
\def\v#1{{\mib #1}}

\def\k{\v{k}}

\def\half{{1\over 2}}

\def\b{\beta}
\def\dl{\delta}
\def\DL{\Delta}
\def\dag{\dagger}
\def\ve{\varepsilon}
\def\pind{\parindent}
\def\hs{\hskip}
\def\vs{\vskip}
\def\hl{\hfil}
\def\ps{\parskip}

\def\hll{\hfill}

\def\ov#1#2{{#1\over#2}}

\def\ua{\uparrow}
\def\da{\downarrow}
\def\ss#1{{\scriptstyle#1}}
\def\mvr{\moveright}
\def\endeq#1{\;${#1}\hll}
\def\Fig#1{Fig.{#1}:\vs-10.5pt}
\def\FPR{\hsize128mm\pind0pt}
\def\math#1{$\,#1\,$}
\def\sub#1{\lower.7ex\hbox{#1}}
\def\sup#1{\lower-.7ex\hbox{#1}}

\def\lsum{{\tenex\char'130}}

\def\lint{{\tenex\char'132}}
\def\mlsum#1#2#3{\lower-2.2ex\hbox to 3em{\hl\lsum\hl}{\lower3ex\hbox to 3em
{\hs-6em{\hl$\ss{#1}$\hl}}}{\lower-3ex\hbox to 3em{\hskip-11.5em{\hl$\ss{#2}$
\hl}}}\hbox{\hskip-6em {$#3$}}}

\def\mlint#1#2#3{\lower-3.5ex\hbox to 1.5em{\hl\lint\hl}{\lower3.5ex\hbox to
3em{\hskip-3.5em{\hl$\ss{#1}$ \hl}}}{\lower-3.5ex\hbox to 3em{\hskip-8.5em{\hl
$\ss{#2}$\hl}}}\hbox{\hskip-5.5em {$#3$}}}


\baselineskip20pt
\parindent12pt
\parskip0pt
\hsize=8cm
\hsize=14cm

\catcode`@=11
\newcount\r@fcount \r@fcount=0
\newcount\r@fcurr
\immediate\newwrite\reffile
\newif\ifr@ffile\r@ffilefalse
\def\w@rnwrite#1{\ifr@ffile\immediate\write\reffile{#1}\fi\message{#1}}

\def\writer@f#1>>{}
\def\referencefile{
\r@ffiletrue\immediate\openout\reffile=ref
  \def\writer@f##1>>{\ifr@ffile\immediate\write\reffile%
    {\noexpand\refis{##1} = \csname r@fnum##1\endcsname = %
     \expandafter\expandafter\expandafter\strip@t\expandafter%
     \meaning\csname r@ftext\csname r@fnum##1\endcsname\endcsname}\fi}%
  \def\strip@t##1>>{}}

\def\citeall#1{\xdef#1##1{#1{\noexpand\cite{##1}}}}
\def\cite#1{\each@rg\citer@nge{#1}} 

\def\each@rg#1#2{{\let\thecsname=#1\expandafter\first@rg#2,\end,}}
\def\first@rg#1,{\thecsname{#1}\apply@rg} 
\def\apply@rg#1,{\ifx\end#1\let\next=\relax
\else,\thecsname{#1}\let\next=\apply@rg\fi\next}

\def\citer@nge#1{\citedor@nge#1-\end-} 
\def\citer@ngeat#1\end-{#1}
\def\citedor@nge#1-#2-{\ifx\end#2\r@featspace#1 
  \else\citel@@p{#1}{#2}\citer@ngeat\fi} 
\def\citel@@p#1#2{\ifnum#1>#2{\errmessage{Reference range #1-#2\space is bad.}%
    \errhelp{If you cite a series of references by the notation M-N, then M and
    N must be integers, and N must be greater than or equal to M.}}\else%
 {\count0=#1\count1=#2\advance\count1
by1\relax\expandafter\r@fcite\the\count0,%
  \loop\advance\count0 by1\relax
    \ifnum\count0<\count1,\expandafter\r@fcite\the\count0,%
  \repeat}\fi}

\def\r@featspace#1#2 {\r@fcite#1#2,} 
\def\r@fcite#1,{\ifuncit@d{#1}
    \newr@f{#1}%
    \expandafter\gdef\csname r@ftext\number\r@fcount\endcsname%
                     {\message{Reference #1 to be supplied.}%
                      \writer@f#1>>#1 to be supplied.\par}%
 \fi%
 \csname r@fnum#1\endcsname}
\def\ifuncit@d#1{\expandafter\ifx\csname r@fnum#1\endcsname\relax}%
\def\newr@f#1{\global\advance\r@fcount by1%
    \expandafter\xdef\csname r@fnum#1\endcsname{\number\r@fcount}}

\let\r@fis=\refis   
\def\refis#1#2#3\par{\ifuncit@d{#1}
   \newr@f{#1}%
   \w@rnwrite{Reference #1=\number\r@fcount\space is not cited up to now.}\fi%
  \expandafter\gdef\csname r@ftext\csname r@fnum#1\endcsname\endcsname%
  {\writer@f#1>>#2#3\par}}

\def\ignoreuncited{
   \def\refis##1##2##3\par{\ifuncit@d{##1}%
     \else\expandafter\gdef\csname r@ftext\csname
r@fnum##1\endcsname\endcsname%
     {\writer@f##1>>##2##3\par}\fi}}

\def\r@ferr{\endreferences\errmessage{I was expecting to see
\noexpand\endreferences before now;  I have inserted it here.}}
\let\r@ferences=\references
\def\references{\r@ferences\def\endmode{\r@ferr\par\endgroup}}

\let\endr@ferences=\endreferences
\def\endreferences{\r@fcurr=0
  {\loop\ifnum\r@fcurr<\r@fcount
    \advance\r@fcurr by 1\relax\expandafter\r@fis\expandafter{\number\r@fcurr}%
    \csname r@ftext\number\r@fcurr\endcsname%
  \repeat}\gdef\r@ferr{}\endr@ferences}


\let\r@fend=\endpaper\gdef\endpaper{\ifr@ffile
\immediate\write16{Cross References written on []\jobname.ref.}\fi\r@fend}

\catcode`@=12

\citeall\refto  

\ignoreuncited
{\vskip 2in}
\parskip16pt
{\Bf
\centerline{Superconductivity of Carriers Doped into }\par
\vskip14pt
\centerline{the Static Charge Density Wave State}\par
\vskip14pt
\centerline{in 2-Dimensional Square Lattice}}\par
{\vskip 1.5cm}
\centerline{\Rm Junichiro Kishine and Hiroshi Namaizawa} \par
{\vskip .5cm}
{\it
\centerline{Institute of Physics, College of Arts and Sciences,} \par
\centerline{University of Tokyo,} \par
\centerline{Komaba 3-8-1, Meguro-ku, Tokyo 153, Japan}} \par\parindent24pt
{\vskip 2cm}
On the purpose of studying the effect of long-range Coulomb-interaction in
strongly correlated electronic systems we bring in as its representative the
nearest-neighbor repulsion ($v$) in addition to the on-site
repulsion ($u$) and shall investigate the possibility of the superconducting
transition of carriers doped into the charge-density wave (CDW) state
expected for $v\,>\,u/4$ in  2-dimensional square lattice.  We shall see
that strongly correlated hopping processes of doped carriers make the systems
superconducting. The favored superconducting phase is of extended s-wave
symmetry, and $T_{c} \sim$100K is shown to easily be attained near the
half-filling.
\vfill
\eject
\leftline{{\Bf \S 1.\quad Introduction}}

\parindent12pt
In studying strong correlation among electrons culminating in high T$_c$
cuprates, heavy fermion systems or organic conductors, one of the most
intricate problems in condensed-matter physics currently, we call for the
Hubbard model and its derivatives like $t$-$J$ model. It is true that we have
learnt much from these models but it is not at all certain whether the learning
survives, even qualitatively, when the neglected long-range Coulomb
interactions are restored. This is mainly because these models are, in spite of
 being simple, not tractable analytically, even in perturbation theory. To say
 the least of it we still do not know what the ground state of the
 two-dimensional Heisenberg antiferromagnet is, the most important starting
 point toward real understanding of high T$_c$ copper oxides as commonly
 believed.
\par
In viewing the status it is worthwhile putting the problem other way around,
namely, taking into account the long-rangeness of Coulomb interaction, to some
extent, from the beginning. The new "Problemstellung" is physically more fit,
especially for cuprates, because of their ionic character. For this purpose
we have proposed \refto{ref1} a model in which the nearest neighbor Coulomb
repulsion, $v$, is brought in as an representative of long-range Coulomb
repulsion, in addition to the on-site repulsion, $u$ and the nearest neighbor
hopping integral, $t$:
$$
H=-t\sum_{<i,j>\sigma}
(c^{\dagger}_{i,\sigma}c_{j,\sigma}+c^{\dagger}_{j,\sigma}c_{i,\sigma})
+u\sum_{i}n_{i,\uparrow}n_{j,\downarrow}
+v\sum_{<i,j>\sigma,\sigma'}n_{i,\sigma}n_{j,\sigma'}\eqno (1-1)
$$\parindent0pt
where $c^{\dagger}_{i,\sigma} (c_{i,\sigma})$ is the creation (annihilation)
operator of an electron of spin $\sigma$ at site $i$, $n_{i,\sigma}$ is the
corresponding number operator with $n_i$ the total electron number at the site
and $<i,j>$ stands for a nearest neighbor combination. We shall call this model
as the $t$--$(u,v)$ model.
\footnote{$^\dagger$}
{\eitrm\baselineskip10pt
We thought it better to avoid calling it as "extended Hubbard model" as
sometimes done, since the same calling is often used for ones, e.g., including
the hopping term to the next nearest neighbors, $\scriptstyle t'$, the one
which may be called as $\scriptstyle (t,t')-u$ model instead.
In the same token the Hubbard model itself might be called as $\scriptstyle t-
u$ model.}
With the model we have discussed \refto{ ref1} the possibility of Cooper-pair
formation in 2-dimensional square lattice  when doped and $v>u/4\gg|t|$. As
shown there, an advantage of the model is that the ground
state is known, in contrast to the conventionally assumed case of
antiferromagnetic (AF) uniform charge distribution ($v<u/4$), to be the
static, nonmagnetic charge-density wave (CDW) state in which lattice sites
are separated into
A-sublattice with two electrons in a spin singlet pair and
B-sublattice with no electron (Fig.1).
The observation is confirmed later
numerically by the quantum Monte Carlo simulation \refto{ref2} and by the
exact diagonalization method \refto{ref3}; Indeed at the absolute zero the
ground state at half filling can be identified as an antiferromagnetic
spin-density wave (SDW) state for $w=u/v>4$ while for $w<4$, as the
CDW state as predicted. Further, for $w<4$, the CDW phase is shown to be
rather rigid against doping \refto{ref2,ref3} even up to  quarter filling
\refto{ref4}. Recently it is also shown \refto{ref5} that the  true ground
state at half filling is the CDW state modified by the condensed charge
fluctuation due to the hopping, just like the ground states of spin ordered
states renormalized by spin waves. The single particle as well as collective
excitaions are found to  be gapped with rather wide openings  (about $8v-u$ for
the former and $7v-u$ for the latter). Thus, as long as \math{|t|\ll v} or
\math{u}, the thermal fluctuations do not disturb the ground state at
temperatures even as high as \math{v/k_B}.\parindent12pt

It is true that the system described by the model for $w<4$ is non magnetic,
charge-density wave ordered and exhibiting a small Fermi surface when doped,
not necessarily in accord with experimental observations up to now, thus
seems not relevant to high $T_c$ cuprates, as it stands. But, as mentioned
above, it has definite advantages, namely the well defined ground state being
rigid against doping and thermal fluctuation, so that it is worth further
theoretical investigation.

As such we would like to pursue the line set by ref.1) and study
in this paper superconductivity of the model when doped. Since the model is
particle-hole
symmetric we shall only consider the hole doping of $N_h=2N-N_e$ holes
where $N_e$ is the total electron number and $2N$ is the number of whole sites.
In \S.2 we present the effective Hamiltonian for the doped holes given in
ref.1) based on the second-order perturbation theory. Elimination of on-site
processes is carried out systematically by introducing pseudo-spin operators
in terms of on-site-avoiding hole operators (construction of the
pseudo-spin operator is done in Appendix A). As a result of the
elimination the Bloch energy of a hole and the hole-hole interaction are
modified. Pairing of holes is favored, especially near the CDW-AF phase
boundary, for the spin-singlet and $s$--wave like state as
in the ordinary BCS mechanism.

Due to the symmetry charactor of the gap function ($\Gamma_4$ of the point
group $D_4$), the corresponding gap equation is brought into a $3\times 3$
matrix form in \S.3 (its elements will explicitly be given in Appendix B).
We shall solve the gap equation iteratively and present
the results for the transition temperature, $T_c$, as a function of hole
concentration, $n_h=N_h/2N$ for various values of $w$: It shows a maximum
around $n_h\sim 0.05$ and can reach a value $k_BT_c\sim J/10$ near
the CDW-AF phase boundary ($w=4$) with $J=t^2/4u$, the exchange integral
adopted by the ordinary Hubbard model, thus $T_c$ can easily be $\sim$ 100K.
We also calculate the superconducting  correlation function and the specific
heat. It is  found that the superconducting correlation length is about
$2.8a_0$ where $a_0$ is the lattice constant of the original square lattice.

Finally in \S.4  we give conclusions and discussions.
\leftline{{\Bf \S.2 Possibility of Cooper Pair Formation}}
\par
Based on ref.1) we shall study the possibility of superconductivity of doped
carriers (holes) in this section.

As noted there, when doped, holes occupy the A-sublattice sites (corresponding
 to Cu$^+$, in case of CuO$_2$ system) while B-sublattice (corresponding to
 Cu$^{+3}$, {\it ditto}) remains intact
unless doped heavily. Hopping integrals and potentials of holes to the second
order in $t^2$ are summarized in terms of $w\,=\,u/v$ and $g\,=\,t^2/v$ in the
following:\parindent0pt

{\it A)Hopping integrals:}\enskip(Fig.2a)\quad  There are two free
hopping integrals
$$
\left\{\eq{
   T_{1}&=-2g/(6-w)\;,\cr
   T_{2}&=T_{1}/2\;, \cr
}\right.\eqno(2-1)
$$
and also two for correlated hopping
\parindent0pt
\footnote{$^\dagger$}
{\eitrm\baselineskip10pt
Correlated hopping mechanism first presented in ref1) is analogious to
the superexchange mechanism of the attractive  interaction between two
localized spins
which leads the system to  the antifferomagnetic ordering. In both mechanism
an intermediate state with higher energy plays an essential role and produces
effective
attraction between two carriers.
}
$$
\left\{\eq{
\bar{T}_{1}=-g\{1/(5-w)+1/(6-w)\}\;,\cr
\bar{T}_{2}=-g/(5-w)\hfill\;.\cr}
\right.\eqno(2-2)
$$
\par
See Fig.2b) for a typical correlated hoppig process.
\par
{\it B)Hole-hole potentials:}\enskip The potential energy $V_m$ for holes of
m-th neighbor apart is
$$
\left\{
\eq{
&V_{1}=-8g\{1/60+1/(5-w)-3/(6-w)+2/(7-w)\}\hskip5mm {\rm for}\enskip m=1\;, \cr
&V_{2}=V_{1}/2\hskip7.2cm {\rm for} \enskip m=2\;,\cr
&V_{m}=0\hskip7.7cm{\rm for}\enskip m\geq 3\;.\cr
}\right.\eqno(2-3)
$$
\par
To the above we must add, at least formally, hopping integrals and potentials
including on-site repulsion, namely, $\bar{T}_1^{(on)},\enskip \bar{T}_2^{(on)
}$ and $V_0$. All of them are of order of $u$ and should properly be
eliminated, as will be seen.\parindent12pt

Taking  all the above into account and performing Fourier-transform
over A-sublattice sites we get  Hamiltonian for the  doped hole
system in the following form:
$$\eq{
{\cal H}\,=\,{\cal H}_0\,+\,{\cal H}_I\;\;,\hskip3cm (2-4{\rm a})\cr
{\cal H}_0\,=\,\sum_{\k,\sigma}
\varepsilon({\k}){b}^{\dagger}_{\k,\sigma}
{b}_{\k,\sigma}\;\;,\hskip3cm (2-4{\rm b})
\cr
{\cal H}_I\,=\,(2N)^{-1}\sum_{\k,\k ',\v {q},\sigma,\sigma '}
{\cal V}_{h}({\k,\k '}){b}^{\dagger }_{\k,\sigma}
b_{\k+\v{q},\sigma}
{b}^{\dagger}_{\k'+\v{q},\sigma'} {b}_{\k',\sigma'}\;\;,\hskip3cm (2-4{\rm c})
\cr
}$$\parindent0pt

where $N$ is the number of A-sublattice sites.  Further
${b}^{\dagger }_{\k,\sigma}$ (${b}_{\k,\sigma}$)
denotes the creation (annihilation) operator of a hole defined by
$$
\left\{
\eq{
&{b}^{\dagger}_{\k,\sigma}\,=\,N^{-1/2}\ds{\sum_{{\v{R}}_{i}\in {\mib A}}}
e^{-i{\k}\cdot{\v{R}}_{i} } c_{i,-\sigma}\;\;,\cr
&{b}_{\k,\sigma}\,=\,N^{-1/2}\ds{\sum_{{\v{R}}_{i}\in {\mib A}}}
e^{i{\k}\cdot{\v{R}}_{i} } c^{\dagger}_{i,-\sigma}\;\;,\cr
}\right. \eqno(2-5)
$$
where $c_{i,-\sigma}$ ($c^{\dagger}_{i,-\sigma}$) is the creation
(anihilation) operator of an electron which appeared in (1--1) and ${\mib A}$
represents the set of all the A-sublattice sites. In (2--4b) $\ve(\k)$ is
the tight-binding Bloch-energy relative to the band bottom
$$\eq{
\ve\,(\k)\,&=-2T_1\,(3-\cosx-\cosy-\cosx\cosy)\cr
&=\ov{W}{4}(3-\cosx-\cosy-\cosx\cosy)\;,\cr
} \eqno(2-6)
$$
where $a=\sqrt{2}\,a_0$ ($\,a_0\,$ is the lattice constant of the
original square-lattice) denotes the lattice constant of the A-sublattice and
\math{W} stands for the band width defined by
$$
W=8|T_1|=16g/(6-w)\;. \eqno(2-7)
$$
Notice that the band bottom lies at \math{6T_1} relative to the energy of
a localized single hole\refto{ref1}.
Finally the hole-hole interaction, ${\cal V}_h$, in (2--4c) can be given by
the aforementioned hopping integrals and potentials as
$$
{\cal V}_{h}({\k,\k'})\,=\,V({\k,\k'})+2\t{T}({\k,\k'})\;,
\eqno(2-8a)
$$
where
$V({\k,\k'})$ and $\t{T}({\k,\k'})$ are respectively given by
$$\eq{
V({\k,\k'})
\,=\,V_{0}+2 V_1 \{\cos(k'_{x}-k_{x})a+\cos(k'_{y}-k_{y})a\}\cr
+4V_{2}\cos(k'_{x}-k_{x})a\cos(k'_{y}-k_{y})a\;,
} \eqno(2-8b)$$
$$
\eq{\t{T}({\k,\k'})\,=\, 2\t{T}^{(on)}_1(\cosx+\cosy+({\k}
\leftrightarrow {\k'}))\cr
+4\t{T}^{(on)}_2(\cosx\cosy+({\k} \leftrightarrow {\k'}))\cr
+4\t{T}_1\{\cos(k_{x}-k'_{x})a\cos k'_{y}a+
\cos k'_{x}a\cos(k_{y}-k'_{y})a+
({\k} \leftrightarrow {\k'})\}\cr
+4\t{T}_{2}(\cosx \cos k'_{y}a+\cosy \cos k'_{x}a)\;.
}\eqno{(2-8c)}
$$
The tilded hopping integrals are defined by subtracting from the correlated
ones the corresponding free hopping integrals, namely, $\t{T}_i=\bar{T}_i-T_i$,
 since the latter are already taken into account in defining the Bloch energy.
 Thus, for example, $\t{T}_1=\t{T}_2=-g\{1/(5-w)-1/(6-w)\}$.
\parindent14pt

In order to discuss the possibility of superconductivity in the doped carrier
system it is convenient to introduce pairing operators defined by
$$
t_a({\k})\,=\,b_{-\k,\sigma}(-i\sigma_2\,\sigma_a\,)_{\sigma\,\sigma'}
b_{\k,\sigma'}/\sqrt{2}\;,\quad a\,=\,0,1,2,3\;,\eqno(2-9a)
$$
where $\sigma_0$ is the $2\times2$ unit matrix and $\sigma_a$ for $a=1,2,3$ are
conventional Pauli-matrices. Needless to say $t_0({\mib k})$ corresponds to
the singlet pairing while ${\mib t(k)}$ governs the triplet pairing.  Further
they satisfy
$$t_0(-{\k})\,=\,t_0({\k})\quad {\rm and}\quad {\mib t}(-{\k})\,=\,-{\mib t}
({\k})\;.\eqno(2-9b)$$

In terms of the pairing operators the reduced Hamiltonian corresponding to
${\cal H}_I$ can be written as
$$\eq{
\hskip2cm{\cal H}_I\,=&\,{\cal H}_{ISP}\,+\,{\cal H}_{ITP}\;,\hskip4.7cm
(2-10{\rm a})\cr
\hskip2cm{\cal H}_{ISP}\,=&\,(2N)^{-1}\,\sum_{\k,\k '}
{\cal V}_{+}({\k,\k '})\,t_0^\dagger({\k}')\,t_0({\k})\;,\hskip2cm (2-10{\rm
b})\cr
\hskip2cm{\cal H}_{ITP}\,=&\,(2N)^{-1}\,\sum_{\k,\k '}
{\cal V}_{-}({\k,\k '})\;{\mib t}^\dagger({\k}')\cdot{\mib t}({\k})\;,\hskip2cm
(2-10{\rm c})\cr
}$$
where ${\cal H}_{ISP}$ and ${\cal H}_{ITP}$ are the reduced interactions
respectively for singlet and triplet pairings and
$$
{\cal V}_{\pm}({\k,\k'})\,=\,({\cal V}({\k,\k'})\pm {\cal V}({\k,-\k'}))/2\;.
\hskip3cm \eqno(2-10d)
$$\parindent14pt

We now concern ourselves with the elimination of "on-site" terms. In the case
of triplet pairing we are automatically free from them owing to the exclusion
principle so that we can use ${\cal H}_{ITP}$ as it stands with ${\cal V}_-
({\k,\k'})$ of (2--10d) given by
$$\eq{
{\cal V}_-({\k,\k'})\,=\,{\cal V}^{(p)}({\k,\k'})\,=\,
2V_1 (\sinx\sinxd +\siny \sinyd)\cr
+4V_2 (
\sinx\cosy\sinxd \cosyd+\cr
\cosx\siny \cosxd\sinyd)
\cr
+8\t{T}_1(\sinx\sinxd\cosyd+\cosxd\siny\sinyd\cr
+\sinx\sinxd\cosy+\cosx\siny\sinyd)\;.\cr
}\eqno(2-11)
$$
For singlet pairing, on the other hand, we have to restrict the Hilbert-space
to the states of no doubly occupied sites.  By so doing we can discard the
terms explicitly pertainig to on-site quantities like
$V_0,\enskip \t{T}_i^{(on)}$.  Further we have to eliminate the on-site
processes inherent in the Fourier-transformed operators.  To do so we replace,
 in a  conventinal manner, the hole operators by
$$
\left\{
\eqalign{
&\tilde{b}^{\dagger}_{\k,\sigma}\,
=\,N^{-1}\,\sum_{\v{R}_{i}\in \mib{A}}e^{-i \k\cdot \v{R}_{i}}
{\tilde{b}}^{\dagger}_{i,\sigma}\;\;,\cr
&\tilde{b}_{\k,\sigma}\,
=\,N^{-1}\,\sum_{\v{R}_{i}\in \mib{A}}
e^{i \v{k}\cdot \v{R}_{i}}
{\tilde{b}}_{i,\sigma}\;\;,\cr
}\right.\eqno(2-12a)
$$
where ${\tilde {b}}^{\dagger}_{i,\sigma}$ and ${\tilde {b}}_{i,\sigma}$ are the
on-site-avoiding operators defined respectively by
$$\left\{
\eqalign{
{\tilde {b}}^{\dagger}_{i,\sigma}&={b}^{\dagger}_{i,\sigma}(1-\bar{n}_
{i,-\sigma})\;\;,\cr
{\tilde {b}}_{i,\sigma}&=(1-\bar{n}_{i,-\sigma}){b}_{i,\sigma}\;\;,\cr}
\right.\eqno(2-12b)
$$
where \math{\bar{n}_{i,\sigma}=b^\dag_{i,\sigma}b_{i,\sigma}=c_{i,-\sigma}
c^\dag_{i,-\sigma}} denotes the number operator of holes at site \math{i} of
spin \math{\sigma}. Then ${\cal H}_0$ of (2--4b) should be replaced by
$${\t{\cal H}}_0\,=\,\sum_{\k,\sigma}
\varepsilon({\k}){\t{b}}^{\dagger}_{\k,\sigma}
{\t{b}}_{\k,\sigma}\;,\hskip3cm \eqno(2-13a)$$
and ${\cal H}_{ISP}$ by
$${\t{\cal H}}_{ISP}\,=\,N^{-1}\,\sum_{\k,\k '}
{\cal V}_{+}({\k,\k
'})\,\t{t}_0^\dagger({\k})\,\t{t}_0({\k}')\;,\hskip3cm\eqno(2-13b)
$$
with $\t{t}_0({\k})$ defined by the on-site-avoiding operators through
(2--9a) and with ${\cal V}_{+}({\k,\k '})$ given as
$$
{\cal V}_{+}({\k,\k'})\,=\,{\cal V}^{(s)}({\k,\k'})+{\cal V}^{(d)}({\k,\k'})\;,
\eqno(2-13c)
$$
where
$$\eq{
{\cal V}^{(s)}({\k,\k'})=
2V_1 (\cosx\cosxd +\cosy \cosyd)\cr
+4V_2 \cosx\cosxd\cosy \cosyd\cr
+8\t{T}_1(\cosx\cosxd\cosyd+\cosy\cosxd\cosyd+\cr
\cosx\cosy\cosxd+\cosx\cosy\cosyd)\cr
+8\t{T}_{2}(\cosx \cosyd+\cosy \cosxd)\cr
-2(6T_1+\ve(\k)+\ve(\k'))\;,
}\eqno(2-13d)$$
and
$$
{\cal V}^{(d)}({\k,\k'})=4V_2 \sinx\sinxd\siny\sinyd\;. \eqno(2-13e)
$$
Note that the last term on the r.h.s. of (2--13d) comes from the fact that in
eliminating the on-site potential term
$$
N^{-1}V_0\sum_{\k}\,t_0^\dagger({\k})\;\sum_{\k'}\,t_0({\k'})\,=\,N^{-1}V_0
\sum_{{\mib R}_i\in {\mib A}}\,b_{i,\uparrow}^\dagger\,b_{i,\downarrow}^\dagger
\,\sum_{{\mib R}_j\in {\mib A}}\,b_{j,\downarrow}\,b_{j,\uparrow}\;,
$$
$V_0$ should be replaced by the Bloch energy of two localized holes,
$2\times6T_1$, and similarly in eliminating the on-site hopping term
the term with $\bar{T}^{(on)}({\k})$ but not with $\t{T}^{(on)}({\k})$
should be taken out, namely
$$
2N^{-1}\sum_{\k}\,\bar{T}^{(on)}({\k})t_0^\dagger({\k})\;\sum_{\k'}\,t_0({\k'})
\;,
$$
thus, in effect $-2(\ve(\k)+6T_1)$ is left\refto{ref1}.

It is rather clear from the definition that the pairing interaction ${\cal V}^
{(p)}$ is of $p$--wave character while ${\cal V}^{(s)}$ and ${\cal V}^{(d)}$
are of $s$-- and $d$--wave character, respectively.  In Figs.4 a) $\sim$ c) we
illustrate the pairing interactions in the first Brillouin zone for $w=u/v=
3.9$. As seen  from them  ${\cal V}^{(s)}$ develops rather a wide and deep
valley around the center of the Brillouin zone,  but the attractive valley of
${\cal V}^{(d)}$ is off the center and shallow ( $|{\cal V}^{(d)}/{\cal V}
^{(s)}|\sim 10^{-1}$ ), so that the $s$--wave like pairing seems favored.
This is mainly because  the correlated hopping processes,
which give rise to large attraction in ${\cal V}^{(s)}$, are absent in
${\cal V}^{(d)}$.  As for the $p-$wave type pairing ${\cal V}^{(p)}$ can be as
large as ${\cal V}^{(s)}$ in magnitude, thanks to the correlated hopping
integral, but the former exhibits alternating valleys and hills around the
center of the Brillouin zone, as can be seen from Figs.4c) so that the
transition temperature higher than that of $s$--wave is unlikely. Thus we
shall study in the following the possibility of superconductivity for the $s$--
wave pairing alone. It is worth noting here that the attractive valley of
${\cal V}^{(s)}$ becomes the deeper and wider the closer \math{w} is to 4, the
CDW--AF boundary: This is mainly due to the correlated hopping terms in (2-11)
which are seen from (2-2) most attractive at \math{w=4} for \math{w\leq 4}.

Now the reduced Hamiltonian to be investigated is
$$\eq{
\t{{\cal H}}^{(s)}\,=\,{\t{\cal H}_0}+N^{-1}\sum_{\k,\k '}
{\cal V}^{(s)}({\k,\k '})\,\t{t}_0^\dagger({\k})\,\t{t}_0({\k}')
\;.}\eqno(2-14a)
$$
For a consistent treatment of the Hamiltonian, the pseudo-spin formalism
\refto{ref6} is most suited. Its application is, however, not immediate since
the hole operators were modified as to avoid on-site double occupation.  As we
shall show in Appendix A we can construct appropriate pseudo--spin
operators from the modified operators, by making use of local mean field
approximation for the commutators among the modified pairing operators,
$\t{b}_{-\k,\downarrow}\t{b}_{\k,\uparrow},\enskip\t{b}_{\k,\uparrow}^\dagger
\t{b}_{-\k,\downarrow}^\dagger$ and
$\t{n}_\sigma({\k})\,=\,\t{b}^{\dagger}_{\k,\sigma}\t{b}_{\k,\sigma}\;,$ as
below:
$$\left\{
\eqalign{
\,\,\t{\tau}^{(z)}({\k})&\,=\,{1\over 2(1-\ov{n_h}{2})}\{\t{n}_
\uparrow({\k})+\t{n}_\downarrow({-\k})-1\}\;\;,\cr
\t{\tau}^{(+)}({\k})&\,=
\,{1\over(1-\ov{n_h}{2})}\t{b}_{\k,\uparrow}^\dagger
\t{b}_{-\k,\downarrow}^\dagger\;\;,\cr
\t{\tau}^{(-)}({\k})&\,=
\,{1\over (1-\ov{n_h}{2})}\t{b}_{-\k,\downarrow}
\t{b}_{\k,\uparrow}\;\;.\cr
}\right.\eqno(2-15)$$
The pseudo-spin operators satisfy  usual $\,SU(2)\,$commutation relations;
$$[\t{\tau}^{(+)}({\k}),\t{\tau}^{(-)}({\k'})]\,=\,2\delta _{\k,\k'}\t{\tau}
^{(z)}({\k})\;,\eqno(2-16a)$$
$$[\t{\tau}^{(z)}({\k}),\t{\tau}^{(\pm)}({\k'})]\,=\,\pm
\delta_{\k,\k'}\t{\tau}^{(\pm)}({\k})\;.\eqno(2-16b)$$
In terms of the pseudo-spin, the reduced Hamiltonian (2--14a) including the
chemical potential term can be rewritten as:
$$\t{\cal H}^{(s)}\,=\,\sum_{\k} (\t{\ve}(\k)-\t{\mu}_h)(2\t{\tau}^{(z)}({\k})
+1)+N^{-1}\,\sum_{\k,\k'}\,\t{\cal V}^{(s)}_{eff}({\k,\k'})\t{\tau}^{(+)}({\k})
\t{\tau}^{(-)}({\k'})\;,\eqno(2-14b)
$$
where \math{\t{\ve}(\k)}, \math{\t{\mu}_h} and \math{\t{\cal V}^{(s)}_{eff}
(\k,\k')} respectively are the Bloch energy, the chemical potential and the
$s$--wave like pairing-interaction modified by the introduction of the
on-site avoiding pseudo-spin and are given by
\pind0pt

\hs2cm$\t{\ve}(\k)=(1-\ov{n_h}{2})\ve(\k)=\t{W}(3-\cosx-\cosy-\cosx\cosy)
\endeq{,}(2-17a)

with \hs4cm $\t{W}=(1-\ov{n_h}{2})W\endeq{,}(2-17b)\pind24pt
$$N_h\,=\,(1-\ov{n_h}{2})\,\sum_\k\,\ov{2}{e^{\b(\t{\ve}(\k)-\t{\mu}_h)}+1}
\enskip,
\eqno(2-18)$$
and
$$\t{\cal V}^{(s)}_{eff}({\k,\k'})\,=\,(1-{n_{h}\over2})^{2}
\enskip{\cal V}^{(s)}(\k,\k')\;. \eqno(2-19)
$$
We can then readily write down the gap equation at a temperature $T$:\pind0pt

\hs3cm$\Delta^{(s)}(\v{k},T)\,=\,-N^{-1}\,$\mlsum{\k'\in \v{C}^<}{}{
\t{\cal V}^{(s)}_{eff}({\k,\k'})\;\DL^{(s)}(\v{k},'T)\,\Theta(\k')}
$\endeq{,}(2--20a)

with\hs2.2cm$\Theta(\k)=\ov{1}{2E({\k})}\tanh({\b E(\k)\over 2})|_{\b=\ov{1}
{k_BT}}\endeq{,}(2--20b)

for the gap defined by
$$ \Delta^{(s)}(\v{k},T)\,=\,-N^{-1}\,\sum_{\k'\in \v{C}^<}
\t{\cal V}^{(s)}_{eff}({\k,\k'})< \t{\tau}^{(x)}({\k})>\;,\eqno(2-20c)$$
where \math{<\,\cdots \,>} denotes the thermal average
$$
<\,\cdots \,>\,=\,{Tr\{ e^{-\b \t{\cal H}^{(s)}}\cdots \}\over
Tr\{ e^{-\b \t{\cal H}^{(s)}}\}}\;,\eqno(2-21)
$$
and \math{\t{\tau}^{(x)}=\half(\t{\tau}^{(+)}+\t{\tau}^{(-)})}. We have taken
the phase convention that \math{< \t{\tau}^{(y)}({\k})>=0} with \math{\t{\tau}^
{(y)}={1\over{2i}}(\t{\tau}^{(+)}-\t{\tau}^{(-)})}. Further $E({\k})$ stands
for the Bogoliubov's quasi-particle energy
$$\left\{
\eq{
\hs5mm E(\k)&=\sqrt{\t{\xi}(\k)^2+\DL^{(s)}(\v{k},T)^2}\;,\cr
{\rm with}\quad\t{\xi}(\k)&=\t{\ve}(\k)-\t{\mu}_h\;.\cr
}\right. \eqno(2-22)
$$
In (2--20a) and (2--20c) $\,\,\v{C}^<\,\,$ stands for the region in the first
Brilluoin zone where the modified Bloch energy is less than an appropriate
cut-off. We shall deal with the cut-off in the next section.

%
%

\vs1cm
\pind24pt
\leftline{{\Bf 3. Properties of the Superconducting Phase}}
\vs5mm
Based on the formalism given in the preceding section we shall solve the gap
equation to determine the transition temperature and the energy gap, then
study the superconducting correlation and the specific heat. We shall omit the
superscript $(s)$ since we restrict ourselves to the $s$--wave like pairing
alone.

{\it The Transiton  Temperature   and  the Energy  Gap:}
Without any loss of generality  we can factorize the gap function as
$$
\DL(\k,T)=d(T)c(\k,T)\;. \eqno(3-1a)
$$
We note that $d(T)$ depends only on
 temperature and \math{c(\v{k},T)} depends both on temperature and wave vector
. The wave-vector part,\math{c(\k,T)}, represents the symmetry of the gap and
directly reflects the lattice symmetry and the anisotropy of the hole-hole
interaction. Then the gap equation (2--20a) can be reduced to
$$
c(\k,T)=-{1\over N}\sum_{\k' \in \v{C}^<}\t{\cal V}_{eff}(\k,\k')
c(\k',T)\,\Theta(\k')\enskip, \eqno(3-2)
$$
where \math{\t{\cal V}_{eff}(\k,\k')} is given by (2--19) and(2--13d).
The transition temperature is determined by the condition
$$
d(T_{c})=0\enskip. \eqno(3-3a)
$$
{}From the hole-hole interaction (2--13d) the wave-vector dependent part,
\math{
c(\k,T)}, should take the following form
\footnote{$^\dag$}
{\eitrm\baselineskip10pt
The form of the gap consists of irreduceble representations of $\ss{\Gamma_{4}}
$ of the point group D$_\ss{4}$: Indeed (3--4) corresponds to the
superposition of the $\ss{s-}$(1) and extended $\ss{s}-$wave solutions
(\math{\ss{\cosx+\cosy}}, \math{\ss{\cosx\cosy}}). There is no
symmetry-related distinction among these
representations.\refto{ref7}}
$$
c(\k,T)=c_0(T)+c_1(T)(\cosx+\cosy)+c_2(T)\cosx\cosy \enskip.
\eqno(3-4)
$$
Before going into the detailed study of the gap equation
, we should note that, although the irreducible representations appearing in
the right hand side of (3-4), \math{1,\;\cosx+\cosy,} and \math{\cosx\cosy},
are independent and othogonal to each other, we cannot decouple (3--2) into
each of these representations due to the presence of the thermal factor,
\math{\Theta(\k')}.  We can only bring the gap equation into the $3 \times 3$
matrix form as
$$
\,\,
\v{M}(T)\,\v{c}(T)=0\quad, \eqno(3-5a)
$$
where
$$
\v{c}(T)=\left(
\matrix{
c_0(T)&\cr
c_1(T)&\cr
c_2(T)&\cr
}\hs-3mm
\right)\quad,\eqno(3-6)
$$
$$\v{M}(T)=\left(
\matrix {
P(T)&Q(T)&R(T)&\cr
S(T)&T(T)&U(T)&\cr
V(T)&W(T)&Z(T)&\cr
}\hs-3mm
\right)\quad.\eqno(3-7)
$$
All the matrix elements appeared in \math{\v{M}(T)} are listed in Appendix B.

It should be noted here that there is an arbitrariness in the factorization,
(3--1a): Indeed an overall factor for \math{c(\k,T)} may be factored out and be
absorbed into \math{d(T)}.
Let us define the gap function at the center of the first Brilloiun zone
($\Gamma$ point) as
$$
\DL_\Gamma(T)=d(T)(c_0(T)+2c_1(T)+c_2(T))\;.\eqno(3-8a)
$$
Now one way for the factorization is to factor out \math{(c_0+2c_1+c_2)} from
\math{c(\k,T)} and use \math{\DL_\Gamma(T)} in place of \math{d(T)}. Instead
 we factor out \math{c_0} from \math{c(\k,T)}, for the sake of numerical
simplicity, and redefine relevant quantities as
$$\hs-9mm\left\{\hs-1mm
\eq{
D(T)&=c_0(T)\,d(T)\;,\cr
C(\k,T)&=c(\k,T)/c_0(T)=1+C_1(T)(\cosx+\cosy)+C_2(T)\cosx\cosy\;,\cr
\hll{\rm with}\;C_i&=c_i/c_0\;,\cr
\DL(\v{k},T)&=D(T)C(\k,T)=D(T)\{1+C_1(T)(\cosx+\cosy)+C_2(T)\cosx\cosy\}\;.\cr
}\hs-5mm\right.\eqno(3-9)
$$
Then \math{\DL_\Gamma(T)} is given by
$$
\DL_\Gamma(T)=D(T)(1+2C_1(T)+C_2(T))\;.\eqno(3-8b)
$$
The gap equation, (3--5a), remains the same formally, but due to the above
mentioned arbitrariness in the vector \math{\v(c)},  should turn out to be
$$
\det{\v{M}(T)}=0\;,\eqno(3-5b)
$$
although \math{\v{M}} contains \math{C_i} in a nonlinear manner through \math
{E(\k)}. The coefficients, \math{C_i}, are self-consistently determined by
$$\left\{
\eq{
C_1&=\ov{RS-PU}{QU-RT}\;,\cr
C_2&=\ov{PT-QS}{QU-RT}\;.\cr
}\right.\eqno(3-5c)
$$
Needless to say the transition temperature is determined from (3--3), or \math
{\DL_\Gamma(T_c)=0}, in other words, by
$$
\det\,\v{M}(T_c)|_{\DL_\Gamma(T_c)=0}\,=0\;,\eqno(3-10)
$$
for a given set of hole concentration, \math{n_{h}}, and \math{w}.

To solve the  gap equation we resort to the iterative  procedure: To begin with
we solve (3--10) and determine the transition temperture, \math{T_c}. Then at
a temperature lower than \math{T_c} we put the zeroth order solution for \math
{C_i(T)} as
$$\left\{\eq{
C^{(0)}_1(T)=C_1(T_c)|_{\DL_\Gamma(T_c)=0}\;,\cr
C^{(0)}_2(T)=C_2(T_c)|_{\DL_\Gamma(T_c)=0}\;,\cr
}\right.\eqno(3-11)
$$
where \math{C_i(T_c)} are determined by (3--5c) by setting \math{T=T_c} and
\math{\DL_\Gamma(T_c)=0}. We next write down the first iteration for \math
{\DL(\k,T)} as
$$
\DL^{(1)}(\k,T)=D^{(1)}(T)\{ 1+C^{(0)}_1(T)(\cosx+\cosy)+C^{(0)}_2(T)
\cosx\cosy \}\enskip, \eqno(3-12)
$$
and solve (3--5b) for \math{D^{(1)}(T)}. By using the solution we calculate
the r.h.s. of (3--5c) to obtain the first iteration to \math{C_i(T)}.
By repeating the procedure for higher iterations until the desired
accuracy is attained we can reach the solution for the gap function at an
arbitrary temperature below \math{T_c}.

We shall now present the numerical results for the transition temperature as
well as the gap function. As for the energy cut-off, \math{\t{\ve}_c},
we take the energy contour corresponding to the Fermi surface at the hole
concentration of \math{n_{hc}=0.25}: This value is chosen because above which
the rigidity of the CDW ground-state for the \math{t-(u,v)} model is shown
demolished, as stated before for CuO$_2$ system\refto{ref4}. Then we have
$$
\t{\ve}_c=0.58\t{W}\;.\eqno(3-13)
$$
We shall call this as the natural cut-off.

Firstly in Fig.5a) we give \math{T_c} as a function of \math{n_h} for sevaral
values of\math{w}. As seen there it increases as \math{\sqrt{n_h}} as \math
{n_h}increases from 0, then develops maximum
at $n_h=0.054$ for $w=4.0$;
at $n_h=0.048$ for $w=3.9$;
at $n_h=0.043$ and $n_h=0.192$ for $w=3.8$, respectively.
It vanishes at \math{n_{hc}^*\simeq n_{hc}} around the cut-off, as
\math{\sqrt{n_{hc}^*-n_h}}. Further, as \math{w} approaches to 4, the curve of
\math{T_c} grows
up in accordance with the observation stated before that the pairing
interaction becomes most attractive at the CDW-AF phase boundary.
The maximum value of the transition temperature reaches $\sim10^{-1}g
$, where the coupling constant $g=t^2/v$ corresponds to the conventional
exchange energy, i.e., \math{4t^2/u},  of the Hubbard-model near $w=4$ and can
be of order of $\sim 10^{3}$K. Hence our mechanism easily explains the
transition temperature as high as \math{\sim 100}K. We have also shown in
Fig.5b) the anisotropy parameters, $C_{1}$ and  $C_{2}$, at \math{T_c} as
functions of \math{n_h} for the same set of values for \math{w}.
Notice that
$$
C_1\;\sim\; C_2\;,\eqno(3-14)
$$
for all the values of \math{w} considered. Further they are rather large at low
 hole concentrations and at the CDW-AF boundary, decrease as \math{n_h}
increases and become small at high concentrations.

As is pointed out just in the above, the transition temperature is dependent
on the choice of the cut-off. To illustrate this we give
\math{T_c} in Fig.6a) and \math{C_i} in Fig.6b) as functions of \math{n_h} at
\math{w=3.9} for several choices of the cut-off. The cut-off dependence can
clearly be seen in Fig.6a). Further, as seen in Fig.6b) the anisotropy is
conspicuous at a low cut-off: This is because the main source of the pairing
interaction is the correlated hopping terms in (2--13d), which are inherently
anisotropic and give strong attraction around the center of the first Brilloiun
 zone, the region most important for a smaller cut-off.

Next we provide the solution to the gap equation at temperatures below \math
{T_c} at the natural cut-off. To be specific we have chosen $w$ to be 3.9.
To obtain an overall view of the gap in \math{\k} space we
plot on Fig.7 \math{\DL(\k,T)} at the absolute zero and \math{n_h=0.05}.
As seen from the figure the gap function is very much alike to the
sign-reversed Bloch energy. This is owing
to the approximate relation (3-14) which is seen holding also at \math{T<T_c}
as shown in Fig.8a). Thanks to this relation we can approximately express the
gap function (3--9) in terms of the modified Bloch energy (2--17a) as
$$
\DL(\k,T)=D(T)\{1+3C_1(T)-4C_1(T)(\ov{\t{\ve}(\k)}{\t{W}})\}\;.\eqno(3-15a)
$$
In other words the gap function has a constant value on a Bloch contour. We
shall thus define from (3--15a) the energy gap, say \math{\DL_F(T)}, of our
model as the gap on the Fermi surface, \math{\t{\ve}(\k)=\t{\mu}_h},  which is
explicitly given by
$$
\DL_F(T)=D(T)\{1+3C_1(T)-4C_1(T)(\ov{\t{\mu}_h}{\t{W}})\}=\DL_\Gamma(T)\{1-
(\ov{4C_1(T)}{1+3C_1(T)})(\ov{\t{\mu}_h}{\t{W}})\}\;.\eqno(3-15b)
$$
In Fig.8b) we have plotted \math{\DL_F(T)} as a function of temperature for
several values of hole concentration. The results for the gaps reveal general
features of the mean field approximation, namely, all of them vanish as
\math{\sqrt{1-{(\ov{T}{T_c})}^2}} at \math{T_c} and develop  wide plateau at
low temperatures. Further in accord with the behavior of \math{T_c} as a
function of \math{n_h} shown in Fig.5a) the gap becomes the widest at
\math{n_h=0.05} at which \math{T_c} is the largest for \math{w=3.9}. Further
we have presented in Fig.8c) \math {\DL_F(0)} and
\math{\DL_\Gamma(0)}, together with \math{\bar{\DL}(0)}, the gap function at
the absolute zero averaged over the 2-dimensional Fermi sea, as functions of
\math{n_h}. Also shown in Fig.8d) are the ratios
$2\DL{_\Gamma}(0)/k_{B}T_c$, $2\DL_{F}(0)/k_{B}T_c$ and $2{\bar\DL(0)}/k_{B}T_c
$ as functions of hole concentration for
$w=3.9$. We can see that the gap on the Fermi contour gives the ratio close to
the BCS value, 3.52.

{}From (3--15b) we may find the condition for the gap on the Fermi surface to
vanish, namely
$$
\ov{\t{\mu}_h}{\t{W}}=\ov{1+3C_1(T)}{4C_1(T)}=\ov{3}{4}+\ov{1}{4C_1(T)}>
\ov{3}{4}\;.\eqno(3-15c)
$$
{}From the above it is clear that the lower bound for \math{\ov{\t{\mu}_h}
{\t{W}}} exceeds the natural cut-off set by (3--13) so that in our model the
gap function never vanishes on the Fermi contour. For a hole concentration of
our interest, i.e., \math{n_h\;\leq\; n_{hc}} the Fermi surface is much
smaller than the contour on which the gap vanishes.

{\it Superconducting Correlation:}
As is already clear so far the gap function shows strong anisotropy reflecting
the lattice symmetry and the anisotropic pairing interaction. It is then of
great interest to know how this anisotropy affects the superconducting
correlation. The superconducting correlation function for the $s$-wave like
pairing can be calculated at \math{T=0} by
$$
\chi_{S}(\v{r})=\ov{1}{2 N{a}^2}\sum_{\k \in \v{C}^<}
\ov{\DL(\v{k},0)\cos k_xx \cos k_yy}{\sqrt{\t{\xi}(\k)^2+\DL(\v{k},0)^2}}\;,
\eqno(3-16)
$$
where $\v{r}=(x,y)$ denotes the relative distance between paired holes in the
A-sublattice coordinates. Recall that the coordinates are rotated by 90 degree
with respect to the coordinates of the original lattice, say $(X,Y)$, so that
the $x=y$--line in the former corresponds to the $Y$--axis in the latter and
the $x$--axis, to the $X=Y$--line. On Fig.9a) we have plotted the correlation
function along the $X=Y$--line, respectively for $n_h=$ 0.03, 0.05, and 0.20
at $w=$3.9. It indicates the superconducting correlation falls rather rapidly
as distance increases and reveals a clear correlation length, \math{\xi},
almost independent of hole concentration, which is estimated to be about
\math{\ov{\xi}{a_0}\simeq2.8} from the plot of maxima of \math{\chi_S}(See
Fig.9b)). We have also found that the superconducting correlation in other
directions is quite similar to the above, thus the superconducting correlation
extends about $\xi\sim 2.8 a_0$ in every direction.

{\it Electronic Specific Heat:}
Electronic specific heat in the superconducting phase
is given by the standard expression;
$$
 C_S=2 {1 \over k_{B} T^2 } \sum_{\k}{e^{\beta {E}_{\k}}
 \over {(e^{\beta {E}_{\k}}+1)}^2}[{E}_{\k}^2
 +\beta \Delta(\v{k},T)
        {d \Delta(\v{k},T) \over d\beta}] \eqno(3-17)
$$
We have evaluated  this quantity  as a function of temperture and
shown the result for \math{w=3.9} and \math{n_h=0.05}
in Fig 10. As a comparison we have also plotted there the heat capacity in the
normal phase,\math{C_N}. The jump ${C_{S}-C_{N}\over C_{N}}$
at \math{T= T_c} becomes $1.67$ for $w=3.9$ and $n_h=0.05$,
a value little larger than the BCS value, $1.43$.
\filbreak
\vskip1cm
\leftline{{\Bf 4. Conclusions and Discussions}}

Based on the 2-dimensional \math{t-(u,v)} model we have shown that the static
, nonmagnetic CDW state, consisting of alternate doubly occupied sites
(A-sublattice) and empty sites (B-sublattice), is eligible for
superconductivity with \math{T_c} as high as 10$^2$K when the system is,
\ps0pt\pind12pt

1) doped near the half-filling and

2) close to the phase boundary to AF.\ps12pt

Main source of pairing interaction between doped carriers, say holes which
necessarily reside on A-sublattice (in case electrons, on B-sublattice), comes
from the correlated hopping caused by consecutive hoppings of the background
electrons via a nearby empty site (e.g., Fig.2b)). We have paid special
attention  to elimination of the on-site processes by making
use of the pseudo-spin formalism. In accord with the elimination, the
Hamiltonian of the doped system has to be modified. We have shown that the
effective hole-hole interaction especially enhances the s-wave type
attraction and results in the highest superconducting-transition temperature
at the CDW-AF phase boundary, $w=4$.

Throughout this paper we have assumed that the background CDW configuration
remains rigid against hole doping for \math{n_h} less than $n_{hc}=0.25$, the
critical dose set by the numerical study for CuO$_2$ system \refto{ref4}.
If more holes are doped, the CDW configuration will make a transition to
a metallic phase.

As noted in the introduction the CDW ground-state is shown \refto{ref5}
modified by the condensed charge fluctuation: There is in effect a fraction of
charges on the originally empty sites, although the CDW state remains
stable even at temperatures as high as \math{v/k_B}. The finding is at the
half-filling, thus it is of due interest to know how doping affects the ground
state and the interaction among doped carriers.  By doing so we can shed light
on the strongly correlated electronic system starting from the localized
configuration, a picture complementary to the conventional ones.

Lastly a few remarks should be made whether our model is relevant to real
superconductors.  CDW configurations similar to ours are known to occur in
substances such as polyacetylene(1D), NbSe$_{2}$(2D) and BaBiO$_{3}$ (3D). In
these systems the CDW instability is driven by the the Fermi-surface
instability. For example  BaBiO$_{3}$ system is a semiconductor with the
3-dimensional CDW-configuration Ba$_{2}$Bi$^{3+}$Bi$^{5+}$O$_{6}$. If Bi is
substituted by Pb the system shows superconductivity at the Pb concentration
of 65\% or higher. The behavior of this compound can be explained in terms of
the modified $t$-$(u,v)$ model, although the Cooper-pair formation seems
mediated by the electron-phonon interaction \refto{ref8}. As for cuprate
high-$T_{c}$ superconductors it is believed that the ground state is
in some AF phase, even doped, and no clue is yet found for any static CDW
configuration.  If, however, \math{w} is very close to 4, where our mechanism
is most efficient for high \math{T_c}, the spin fluctuation might well emerge
and play some role in the ground state as well as in the superconductive phase,
since the boundary to AF phase is immediate. This possibility is to be
investigated as a future problem together with the charge fluctuation
mentioned above.
\vs10mm

\leftline{{\Bf Acknowledgement}}

One of the authors (H. N.) would like to express his sincere thanks to
Professor Y. Mizuno from Science University of Tokyo for his valuable
discussions and continual interests in the early stage of this work.
\filbreak
\beginsection {Appendix A}\pind12pt

We shall construct the pseudo-spin operator in terms of the on-site avoiding
creation and annihilation operators of (2--12b). First we define \math{\tau}-
operators by
$$\left\{
\eq{
\tau^{(+)}({\k})&=
{\tilde{b}}^{\dagger}_{\k,\uparrow}
{\tilde{b}}^{\dagger}_{-\k,\downarrow}
=\ds{N^{-1}\sum_{\mib{R}_{i},\mib{R}_{j}\in \mib{A}}}
e^{i\k\cdot (\mib{R}_{i}-\mib{R}_{j})}
{\tilde{b}}^{\dagger}_{i,\uparrow}
{\tilde{b}}^{\dagger}_{j,\downarrow}\enskip,\cr
\tau^{(-)}({\k})&=
{{\tilde b}}_{-\k,\downarrow}
{{\tilde b}}_{\k,\uparrow}
=\ds{{N^{-1}}\sum_{\mib{R}_{i},\mib{R}_{j}\in \mib{A}}}
e^{-i\k\cdot ({\mib R}_{i}-{\mib R}_{j})}
{\tilde{b}}_{j,\downarrow}
{\tilde{b}}_{i,\uparrow}\enskip.\cr
}\right.\eqno({\rm{A}}-1)
$$
\hb
Then we calculate the commutation relation of \math{\tau}-operators, i.e.,
$$[\tau^{(+)}({\k})
,\tau^{(-)}({\k}')]
=\ds{\sum_{{\mib R}_{i},{\mib R}_{j},
{\mib R}_{l},{\mib R}_{m}\in {\mib A}}}
e^{i \k \cdot ({\mib R}_{i}-{\mib R}_{j})
-i\k'({\mib R}_{m}-{\mib R}_{l})}
[
{\tilde{b}}^{\dagger}_{i,\uparrow}
{\tilde{b}}^{\dagger}_{j,\downarrow}
,{\tilde{b}}_{l,\downarrow}
\tilde{b}_{m,\uparrow}]\enskip.
$$
\hb
After some algebra we have for the commutator on the right hand side above
$$\eqalignno{
[{\tilde{b}}^{\dagger}_{i,\uparrow}
{\tilde{b}}^{\dagger}_{j,\downarrow}
,{\tilde{b}}_{l,,\downarrow}
{\tilde{b}}_{m,\uparrow}] =&
\delta_{jl}(1-\bar{n}_{j,\uparrow})
\t{b}^{\dagger}_{i,\uparrow}
\t{b}_{m,\uparrow}
-\delta_{jl}\delta_{im}(1-\delta_{il})\t{b}^{\dagger}_{i,\uparrow}
\t{b}_{l,\uparrow}\cr
&+\delta_{im}(1-\bar{n}_{i,\uparrow})\t{b}^{\dagger}_{j,\downarrow}
\t{b}_{l,\downarrow}
-\delta_{im}\delta_{il}(1-\delta_{ij})\t{b}^{\dagger}_{j,\downarrow}
\t{b}_{i,\downarrow}\cr
&-\delta_{im}\delta_{jl}(1-\delta_{ij})(1-\bar{n}_{i,\da})(1-\bar{n}_{j,\ua})\cr
&+\delta_{ij}\delta_{il}\t{b}^{\dagger}_{i,\uparrow}\t{b}_{m,\uparrow}+
\delta_{jm}\delta_{jl}\t{b}^{\dagger}_{i,\uparrow}\t{b}_{j,\uparrow}-
\delta_{jm}\t{b}^{\dagger}_{i,\uparrow}\t{b}_{l,\downarrow}
\t{b}^{\dagger}_{j,\downarrow}\t{b}_{j,\uparrow}\cr
&-\delta_{il}\t{b}^{\dagger}_{i,\uparrow}\t{b}_{i,\downarrow}
           \t{b}^{\dagger}_{j,\downarrow}\t{b}_{m,\uparrow}
+\delta_{jm}\delta_{il}(1-\delta_{ij})\t{b}^{\dagger}_{i,\ua}\t{b}_{i,\da}
           \t{b}^{\dagger}_{j,\da}\t{b}_{j,\ua}\enskip.\cr
}$$
\hb
To the above we apply the local mean-field approximation for the number
operators of holes, i.e.,
$$
\bar{n}_{i\sigma}\; \to\;
<\bar{n}_{i\sigma}>=\ds{{n_{h}\over 2}} \enskip.
$$
\hb
Further, since the system is nonmagnetic, we can neglect terms which make spins
flip together with terms ending up with \math{O(1/N)} to the main terms when
Fourier transformed.  Thus we obtain
$$
[\tau^{(+)}({\k})
,\tau^{(-)}({\k'})]
=2\dl_{\k\k'}(1-{n_{h}\over 2})\tau^{(z)}(\k)\enskip,
\eqno({\rm{A}}-2)
$$
where
$$
\tau^{(z)}(\k)=\half(\t{n}_\ua(\k)+\t{n}_\da(-\k)-1)\enskip,\enskip{\rm with }
\enskip\t{n}_\sigma(\k)=\t{b}^\dag_{\k,\sigma}\t{b}_{\k,\sigma}\enskip.
\eqno({\rm{A}}-3)
$$
In a similar manner we calculate commutators of \math{\tau^{(z)}} with
\math{\tau^{(\pm)}} to get
$$
[\tau^{(\pm)}({\k}),\tau^{(z)}(\k')]=\pm\dl_{\k,\k'}(1-{{n_h} \over 2})
\tau^{(\pm)}({\k})\enskip.
\eqno({\rm{A}}-4)
$$
\hb
Therefore if we define the pseudo-spin operators, ${\t{\tau}}$, by
$$
{\tilde{\tau}}^{(z)}({\k})
=\half\ov{1}{(1-\ov{n_{h}}{2})}(\t{n}_\ua(\k)+\t{n}_\da(-\k)-1)
=\ov{1}{(1-\ov{n_h}{2})}\tau^{(z)}(\k)\enskip,
\eqno({\rm{A}}-5)
$$
and
$$
{\tilde{\tau}}^{(\pm)}({\k})=\ov{1}{(1-\ov{n_h}{2})}
\left\{\matrix{
{\t{b} }^{\dag}_{\k ,\uparrow}
{\t{b} }^{\dag}_{-\k ,\downarrow}\cr
{\t{b} }_{-\k ,\downarrow}
{\t{b} }_{\k ,\uparrow}\cr
}\right\}=\ov{1}{(1-\ov{n_h}{2})}\tau^{(\pm)}(\k)\enskip,
\eqno({\rm{A}}-6)
$$
then we can show from (A-2,4) that the pseudo-spin operators  satisfy the
$SU(2)$ commutation relations, (2-16a,b).
\filbreak
{\Bf Appendix B}\pind12pt

We shall provide the matrix elements appeared in the equation (3--7). By
inserting  the effective interaction (2--13d) and the gap function (3--4) into
the gap equation (3--2) we can obtain them as follow:
$$\eqalign{
&P(T)=1+4(1-{n_{h}\over 2})^{2}{ T}_{1}(3A(T)-B(T)-C(T))\enskip,\cr
&Q(T)=4(1-{n_{h}\over 2})^{2}{T}_{1}(3B(T)-D(T)-E(T))\enskip,\cr
&R(T)=4(1-{n_{h}\over 2})^{2}{ T}_{1}(3C(T)-E(T)-F(T))\enskip,\cr
&S(T)=-(1-{n_{h}\over 2})^{2}\{V_{1}B(T)+4{\tilde
T}_{1}(2C(T)+B(T))+4T_{1}A(T)\}\enskip,\cr
&T(T)=1-(1-{n_{h}\over 2})^{2}
\{V_{1}D(T)+4{\tilde T}_{1}(2E(T)+D(T))+4T_{1}B(T)\}\enskip,\cr
&U(T)
=-(1-{n_{h}\over 2})^{2}
\{V_{1}E(T)+4{\tilde T}_{1}(2F(T)+E(T))+4T_{1}C(T)\}\enskip,\cr
&V(T)
=-(1-{n_{h}\over 2})^{2}
\{2V_{1}C(T)+8{\tilde T}_{1}B(T)+4T_{1}A(T)\}\enskip,\cr
&W(T)
=-(1-{n_{h}\over 2})^{2}
\{2V_{1}E(T)+8{\tilde T}_{1}D(T)+4T_{1}B(T)\}\enskip,\cr
&Z(T)
=1-(1-{n_{h}\over 2})^{2}
\{2V_{1}F(T)+8{\tilde T}_{1}E(T)+4T_{1}C(T)\}\enskip,\cr
}
$$
where
$$\eqalign{
&A(T)={\overline 1}\enskip,\cr
&B(T)= {\overline {\cosx+\cosy}}  \enskip,\cr
&C(T)= {\overline {\cosx\cosy}}  \enskip,\cr
&D(T)= {\overline {(\cosx+\cosy)^2}}  \enskip,\cr
&E(T)= {\overline {\cosx\cosy(\cosx+\cosy)}}  \enskip,\cr
&F(T)= {\overline {\cos^2{k_x}\cos^2{k_y}}} \enskip.\cr
}
$$
\hb
In the above $\,\,\overline{\cdots \cdots}\,\,\,$ denotes
$$
\,\,\overline{\cdots \cdots}
=-{1\over N}
\displaystyle{
\sum_{\k \in \v{C}^<}
(\cdots \cdots){\tanh{1\over 2}  \beta E ({\k})
\over 2 E ({\k})}
}\enskip.
$$
\filbreak
\refis{ref1} Y. Mizuno and H. Namaizawa \jrnl{Prog. Theor.
Phys.}{80}{1988}{353}.

\refis{ref2} Y. Zhang and J. Callaway \jrnl{Phys.  Rev.}{B39}{1989}{9397}.

\refis{ref3} J. Callaway, D.P. Chen, D.G. Kanhere,
and Quiming Li\jrnl{Phys. Rev.}{B42}{1990}{465}.

\refis{ref4} Y. Ohta, K. Tsutsui, W. Koshibae and S. Maekawa, Nagoya Univ.
Preprint(1993).

\refis{ref5} T. Kawana and H. Namaizawa, in preparation.

\refis{ref6} P.W. Anderson \jrnl{Phys. Rev.}{112}{1958}{1900}.

\refis{ref7} M. Sigrist and T.M. Rice \jrnl{Z. Phys}{B68}{1987}{9}.

\refis{ref8} T.M. Rice and L. Sneddon; \jrnl{Phys. Rev. Lett.}{47}{1981}{689}.

\references

\endreferences

{\bf{Figure Captions}}\pind0pt

\Fig{ 1}
\mvr1.2cm
\vbox{\FPR The static CDW configuration at the half-filling having alternate
doubly occupied sites (A-sublattice) and empty sites(B-sublattice). Arrows
stand for electrons of corresponding direction, and \math{a_0} and \math{a=
\sqrt{2}a_0} respectively are the lattice costants of the original square
lattice and A- or B-sublattice.}
\vs3mm
\Fig{ 2}
\mvr1.2cm
\vbox{\FPR a) Classification of the hopping integrals. The hole to hop is
marked by a circle, while the doubly encircled hole is a spectator; b) An
example of processes leading to the correlated hopping integral, \math
{\t{T}_1}. The dark arrows are for background electrons and the hatched ones
for holes. By consecutive hoppings of a background electron (No.3 in the
figure) via a neighboring empty site, a hole (No.1) in the initial
configuration (\math{|\,i>}) is effectively moved to the final one
(\math{|\,f>}) through the intermediate one (\math{|\,m>}).}
\vs3mm
\Fig{ 3}
\mvr1.2cm
\vbox{\FPR a)The contour of the Bloch energy of doped carriers,
$\varepsilon({\mib k})$, in the first Brillouin zone for $w=3.9$; b) The
chemical potential, $\mu_{h}$, of doped carriers for $w=3.9$.
Energy is in unit of the band width, $W=16g/(6-w)=7.6g$.}
\vs3mm
\Fig{ 4}
\mvr1.2cm
\vbox{\FPR The pairing interactions in the first Brillouin zone along
$k_{x}=-k_{x}'$ and $k_{y}=-k_{y}'$ for\math{w=3.9} respectively: a) of
\math{s-}wave type, \math{{\cal{V}}^{(s)}(\k,\k')}; b) of \math{p-}wave type,
\math{{\cal{V}}^{(p)}(\k,\k')} and c) of \math{d-}wave type, \math{{\cal{V}}
^{(d)}(\k,\k')}. Energy is in unit of the band width, $W=16g/(6-w)=7.6g$.
Notice difference in scale  between  c) and  a) or b).}
\eject
\null\vs-1pt
\hb
\Fig{ 5}
\mvr1.2cm
\vbox{\FPR a)The transition temperature, $T_{c}$, devided by \math{g=\ov{t^2}
{v}} and b) the gap-anisotropy
parameters, $C_{1}$ and $C_{2}$,  as functions of hole
concentration, n$_{h}$, respectively for $w=$4.0, 3.9 and 3.8.  The
energy cut-off corresponds to the chemical potential at the natural cut-off
for the hole concentration, $n_{hc}=0.25$.}
\vs3mm
\Fig{ 6}
\mvr1.2cm
\vbox{\FPR a) The transition temperature in unit of \math{g} and b) the
gap-anisotropy parameters,  at \math{w=3.9} for various values of the energy
cut-off set by
the chemical potential corresponding to the cut-off in hole concentration,
$n_{c}$.}
\vs3mm
\Fig{ 7}
\mvr1.2cm
\vbox{\FPR The gap function, $\Delta(\k,T)$, devided by \math{g}
in the first Brillouin zone at $T=0$ for $w=3.9,n_{h}=0.05$ and the natural
cut-off.}
\vs3mm
\Fig{ 8}
\mvr1.2cm
\vbox{\FPR a) The gap-anisotropy parameters and b)the gap on the Fermi
contour, \math{\DL_F(T)} devided by \math{g}, as functions of temperature for
$n_{h}$=0.03 ,0.05 and 0.20. Also shown are,
c) \math {\ov{\DL_F(0)}{g}}, \math{\ov{\DL_\Gamma(0)}{g}} and \math
{\ov{\bar{\DL}(0)}{g}}, and d) $\ov{2\DL{_\Gamma}(0)}
{k_{B}T_c}$, $\ov{2\DL_{F}(0)}{k_{B}T_c}$ and $\ov{2{\bar\DL(0)}}{k_{B}T_c}$,
respectively as functions of hole concentration, \math{n_h}.  All of them are
calculated at \math{w=\ov{u}{v}=3.9} and the natural cut-off.}
\vs3mm
\Fig{ 9}
\mvr1.2cm
\vbox{\FPR a) The superconducting correlation function and b) the logarithm of
its local maxima along the line X=Y for n$_{h}$=0.03, 0.05 and 0.20 at
\math{w}=3.9}
\vs3mm
\Fig{10}
\mvr1.2cm
\vbox{\FPR The specific heat in the superconducting (\math{C_S}) and normal
(\math{C_N}) phases normalized by \math{C_S(T_c)} for $w=3.9$ and $n_{h}=0.05$
at the natural cut-off.}
\filbreak
\end